\documentstyle{proceed}

\title[Modern models of interstellar dust]{Modern models of interstellar dust}

\author[Zubko]{V.G. Zubko$^{1,2}$\email{zubko@phquasar.technion.ac.il}}

\institute{$^1$Department of Physics, Technion -- Israel Institute of Technology,
               Haifa 32000, Israel \\
           $^2$On leave from the Main Astronomical Observatory, NAS, Kiev,
               Ukraine}

\begin{document}
\maketitle

\abstract{ The new observational data on interstellar dust, obtained
  during the past few years, in particular, revised ISM elemental abundances,
  the polarization curve from the far UV to the near IR, spectra of dust
  infrared emission, extended red emission (ERE), are in serious conflict with
  existing dust models. We proposed here to use the regularization approach
  as a base for developing a modern dust model, which attempts to
  self-consistently explain most of the available observational data.
}

\section{Why is dust important for astronomy?}

Dust plays an important role in various cosmic environments:
circumstellar shells, star-forming regions, protoplanetary nebulae,
interstellar and intergalactic medium (for a review see Dorschner \& Henning
\cite{dh95}). The dust grains absorb, scatter and re-emit
the electromagnetic radiation from various cosmic objects.
The dust grains contribute to the thermal equilibrium of interstellar matter.
Various chemical and physical processes, e.g. formation of molecules,
take place on the surfaces of dust grains. Thus, the knowledge of
the properties of cosmic dust is important for: 1)~the correct interpretation
of the observed spectra of objects seen through intervening dust and
the observations of radiation emitted from, or scattered by dust and
2) modeling of objects where dust is a major constituent.

We may mention a few dust models which were most popular until recently.
One of them is the so-called MRN model (Mathis, Rumple \& Nordsieck \cite{mrn}):
a mixture of spherical bare graphite and silicate grains, having
a power law size distribution $f(a) \sim a^{-3.5}$ over the range of sizes
0.005 -- 0.25 $\mu$m. This model was further developed
by Draine \& Lee (\cite{dl84}), Mathis \& Whiffen (\cite{mw89}),
Kim, Martin \& Hendry (\cite{kmh}), Mathis (\cite{mathis96}).
Another dust model was proposed by Hong \& Greenberg (\cite{hg78}):
small bare graphite grains (for explanation of the 217.5 nm hump) and
silicate grains covered by the organic refractory mantles produced by
UV photolysis of dirty ices which are formed on the grains in molecular clouds.
It was assumed that such complicated silicate grains obey some exponential
cubic size distribution. This model was further developed by
Greenberg (\cite{greenberg89}), Greenberg \& Li (\cite{gl96}),
Li \& Greenberg (\cite{lg97}). Both above-mentioned models were in rather
good agreementwith the mean Galactic extinction and polarization laws,
provided the cosmic elemental abundances are equal to solar abundances.

\section{New observational data on cosmic dust}

During the past few years considerable progress has been achieved
in obtaining new observational data on interstellar dust.
Here are some of the most important findings:
\begin{enumerate}
\item
It became clear that the chemical compositions of the general
interstellar medium and the Sun are not the same, as was previously
assumed: the cosmic elemental abundances appear to be 60--70
per cent of the solar abundances (Snow \& Witt~\cite{sw96}).
\item
The analysis of high signal-to-noise spectra of interstellar
carbon and oxygen in a few directions obtained with the Goddard
High Resolution Spectrograph aboard the {\em Hubble Space Telescope},
produced the first reliable estimates of the mean gas-phase abundances of
interstellar carbon: 10$^6$C/H$\approx$140$\pm$20 and oxygen
10$^6$O/H$\approx$310$\pm$20 (Cardelli et al. \cite{car96}).
\item
It was found that the dust albedo in the near-IR, 0.6--0.8,
considerably exceeds the albedo predicted by the standard dust
models (Witt et al. \cite{witt94}; Lehtinen \& Mattila \cite{lm96}).
Recently, Witt et al. (\cite{wfs97}) derived estimates of both albedo
and asymmetry parameter of dust in the diffuse medium in the far-UV,
which are generally consistent with expectations.
\item
Far-UV polarization observations have been performed
(Clayton et al. \cite{clayton92}; Somerville et al. \cite{somerville94};
Clayton et al. \cite{clayton95}), and produced the wavelength-dependent
interstellar polarization curves from the far-UV down to the near-IR.
\item
The infrared emission from the diffuse interstellar medium
in the range 3.5--1000 $\mu$m has been measured for the first time with
{\em COBE}/DIRBE and FIRAS (Bernard et al. \cite{bernard94};
Dwek et al. \cite{dwek97}; Sodroski et al. \cite{sodroski97};
Lagache et al. \cite{lagache98}).
\item
The so-called extended red emission (ERE) in the 500--800 nm spectral
range has been detected in the general interstellar medium
(Gordon et al. \cite{gordon98}; Szomoru \& Guhathakurta \cite{sg98}).
Witt et al. (\cite{witt98}) and Ledoux et al. (\cite{ledoux98}) independently
proposed that silicon nanoparticles might be the source of the ERE.
\end{enumerate}

\section{Previous work on modeling of interstellar dust}

With the wealth of new data it became clear that practically all
dust models, proposed before 1995 (Dorschner \& Henning \cite{dh95}),
face a serious crisis in explaining the new observational data.
No attempts were undertaken so far to build a dust model which
would accomodate most of the data recently acquired.
However, some important theoretical steps have been done
towards such the model. In particular, Mathis (\cite{mathis96}),
Zubko et al. (\cite{z96}, \cite{z98}) and Li \& Greenberg (\cite{lg97})
proposed more sophisticated dust models on the base of composite,
core-mantle and multilayer grains with the main goal to explain
the interstellar extinction using the revised elemental abundances.
Li \& Greenberg (\cite{lg97}) have made an attempt to analyse
the interstellar extinction and polarization, taking into account
the new chemical abundances. Dwek (\cite{d97}) has tried to model
the extinction and infrared emission together.
Note that the approaches used by Mathis (\cite{mathis96}), Dwek (\cite{d97})
and Li \& Greenberg (\cite{lg97}) were based on the predefined grain-size
distributions: power law or Gaussian and, therefore, cannot be
considered as leading to the best solution. Kim et al. (1994) and
Kim \& Martin (\cite{km94}, \cite{km95}, \cite{km96}) have modelled
the interstellar extinction and polarization using the Maximum Enthropy
Method (MEM). However, the authors treated separately the extinction and
the polarization rather than self-consistently. In addition, as was
noted by Zubko et al. (\cite{z96}), the MEM cannot produce a {\em unique}
solution, because some default solution should be specified prior to
the calculations. As a rule, the default is not known a priori and this
introduces some uncertainty into the final result.

Recently, Zubko (\cite{z97}) has proposed the regularization approach
for modeling of interstellar dust. To the best of our knowledge,
it is the only tool which is capable of deriving an optimum and unique
grain-size distribution in a {\em general} form for any predefined mixture
of model grains. In addition, the regularization approach
directly takes into account the mathematical nature of the problem,
being an ill-posed inverse problem. A few years ago the author started
to develop a computer program based on the regularization approach.
In its present state, this program is capable of building a model of
interstellar dust by a self-consistent analysis of an extinction curve,
scattering properties: albedo and asymmetry parameter,
elemental abundancies and mass fraction constraints.
Currently, the program works with spherical grains which may be
homogeneous (their optical properties are calculated with the standard Mie
approach), multilayer (author's original approach) or of composite
structure (EMT/Mie approach). The first models of both circumstellar and
interstellar dust derived with the program appear to be very promising
(Zubko \cite{z97}; Zubko et al. \cite{z96}, \cite{z98}).

\section{What should a modern dust model fulfill?}

In our view, an ideal model of interstellar dust, which could be
constructed in near future, should fulfill the following requirements:

\begin{enumerate}
  \item
  it should be {\em a unified dust model}, that is, it should accomodate
  most existing observational data on extinction, polarization,
  elemental abundances and so on.
  \item
  it should use dust particles {\em of more realistic structure}:
  composites, fractal, core-mantle and multilayer grains, both spherical
  and non-spherical;
  \item
  it should allow {\em more grain species}, and use the optical
  properties, preferably based on laboratory measurements of cosmic
  analogue grains;
  \item
  it should handle {\em the grain-size distributions in the most
  general form} rather than in some prescribed form, e.g. a power law or
  a Gaussian; this approach should provide an optimum solution.
\end{enumerate}

The best-fit model will be derived using the regularization approach,
which in addition to the already mentioned advantages, will allow one
to easily include other observational constraints in the future.
One of the by-products of this work will be a well-documented program for generating a self-consistent
model of interstellar dust, which will be available upon request.



\acknowledgements{
The report was prepared during my stay in Technion as a Visiting Scientist.
During my work, I benefited from many stimulating discussions with Ari Laor.
This research was supported by a grant from the Israel Science Foundation.
}

\end{document}